# Role of local structural distortions on the origin of *j=1/2* pseudo-spin state in sodium iridate

Priyanka Yadav[1], Sumit Sarkar[1], Manju Sharma[2], Rajamani Raghunathan[1,*], Ram Janay Choudhary[1,*], D.M. Phase[1]

[1]UGC-DAE Consortium for Scientific Research, DAVV Campus, Khandwa Road, Indore – 452001, Madhya Pradesh., India

[2]School of Chemistry, University of Hyderabad, Prof. C.R. Rao Road, Gachibowli, Hyderabad – 500046, Telangana, India.

[*]Corresponding authors: rajamani@csr.res.in, ram@csr.res.in

## *Abstract*

$Na_2IrO_3$ (NIO) is known to be a spin-orbit (SO) driven *j=1/2* pseudo-spin Mott-Hubbard (M-H) insulator. However, the microscopic origin of the pseudo-spin state and the role of local structural distortions have not been clearly understood. Using a combination of theoretical calculations and x-ray spectroscopy, we show that the energetics in the vicinity of Fermi level ($E_F$) is governed by SO interactions, electron correlation and local octahedral distortions. Contrary to the earlier understanding, here we show that the *j=3/2* and *1/2* pseudo-spin states have admixture of both $t_{2g}$ and $e_g$ characters due to local structural distortion. Reduction of local octahedral symmetry also enables Ir $5d$ – O $2p$ hybridization around the $E_F$ resulting in a M-H insulator with enhanced charge transfer character. The possibility of Slater insulator phase is also ruled out by a combination of absence of room temperature DoS in valence band spectra, calculated moments and temperature dependent magnetization measurements.

## *1. Introduction*

Transition metal oxides (TMOs) containing $3d$ transition metal (TM) ions have been studied immensely over the past several decades for its interesting properties including superconductivity, ferroelectricity and magneto-resistance [1-4]. There is a recent spurt in the study of $4d$ and $5d$ TMOs like $Sr_2RuO_4$, $Sr_2IrO_4$, $Na_2IrO_3$ (NIO) and $Cd_2Os_2O_7$, due to their non-trivial electronic structures and exotic quantum properties like Weyl/Dirac semi metal, topological insulator, Kitaev spin liquids and spin-orbit assisted Mott-Hubbard insulator, owing to the presence of strong spin-orbit interaction (SOI) [5-8]. NIO is an interesting $5d$ TMO containing edge-shared $IrO_6$ octahedra arranged in the form of a two-dimensional honeycomb lattice [9,10]. It is believed that the strong SOI of iridium 5$d$ splits the $t_{2g}$ manifold into a lower $j=3/2$ quartet and an upper $j=1/2$ doublet. The doublet further splits into lower and upper Hubbard bands in presence of weak electron correlations ($U$). The local edge-sharing geometry leads to destructive interference of electrons hopping in the clockwise and counter-clockwise directions resulting in a net zero isotropic super-exchange between neighbouring $j=1/2$ pseudo-spins. An Ising type Kitaev super-exchange of the $j=1/2$ pseudo-spins, which stems from the Hund's direct exchange interaction in a multi-orbital Hubbard model gives rise to magnetic frustration, which is critical to realize quantum spin liquid phase for quantum computer applications [11-13].

The electronic structure of NIO has been a topic of constant debate. Preliminary models supported by angle resolved photoemission spectroscopy (ARPES), angle integrated photoemission spectroscopy (AIPES) and optical spectroscopy experiments along with first principles calculations, considered NIO as a spin-orbit (SO) assisted $j=1/2$ Mott insulator arising purely due to SO interactions of the crystal field split $t_{2g}$ states, which is robust even under high pressure [14-19]. But a density functional theory (DFT) calculation using hybrid functional, HSE06 suggested that NIO is a Slater insulator [20]. A previous polarization dependent x-ray linear dichroism study showed that the unoccupied $j=1/2$ state near $E_F$ is admixed with the $j=3/2$ state. Further, the splitting of the $e_g$ state has been attributed to the non-local interactions [21]. But, local structural distortions like the octahedral tilt and trigonal distortion can also lower the ideal cubic symmetry ($O_h$) and are common in non-cubic TMOs [22-25]. Such distortions can lead to energy level splitting and mixing of the spin-orbit entangled pseudo-spin states resulting in deviation from the ideal $j=3/2$ and $1/2$ behaviour, which is detrimental to the realization of spin liquid state. The role of such structural distortion on the electronic structure and its effect on the $j=1/2$ pseudo-spin state have not been explored so far. The larger spatial extent of 5$d$ orbital of iridium compared to 3$d$ TMs can lead to stronger hybridization with O 2$p$ orbitals and enhance the covalency of Ir-O bonds, which should also have implications on its electronic structure. These conundrums regarding electronic state of NIO prompted us to explore it broadly using first-principles calculations and experiments. Contrary to the earlier hypothesis, here we show that the $j=1/2$ pseudo-spin emerges from an admixture of $t_{2g}$ and $e_g$ manifolds, due to local octahedral distortion. We further show that the insulating state of NIO is intermediate between M-H and CT insulators due to hybridization between Ir 5$d$ and O 2$p$ states near $E_F$ and is not a Slater insulator.

## 2. Methodology

The energetics and electronic structures of NIO were calculated within the framework of DFT by employing Vienna *Ab-initio* Simulation Package (VASP) using projector augmented wave (PAW) potential [26-29]. Crystal structures and magnetic properties of NIO are known to be well described within a DFT+*U* formalism by considering $U = 1.7$ eV and $J = 0.6$ eV [23,30,31] including SO and the same parameters were used in our calculations. Plane waves upto 600 eV energy were chosen to describe the valence electrons. Gamma centered *k*-mesh of grid size 6×4×6 and 8×4×8 were taken during self-consistent field method (SCF) and the density of states (DOS) calculations respectively. An energy tolerance of $10^{-8}$ eV between two consecutive steps was chosen for electronic minimization. The starting structures are relaxed with respect to volume and ion positions for each of the four unique magnetic configurations (*Figure 1*). Ionic relaxations were done until the Hellman-Feynman forces becomes less than 5 meV/Å corresponding to an energy convergence of less than $10^{-4}$ eV between two successive ionic steps. Calculations were done with and without spin-orbit interactions.

Polycrystalline sample of $Na_2IrO_3$ was synthesized using solid state route. Initial stoichiometric precursors of $Na_2CO_3$ (10% excess) and $IrO_2$ with purity 99.99% were mixed homogeneously. In order to avoid the deficiency of Na due to its notorious volatile nature, the sample was covered with lid during initial heating. The resulting mixture was once again heated at 900°C for 48 hours and finally pelletized and sintered at 900°C for 48 hours [9], with intermediate grindings for homogenization. Bruker D2 phaser Cu $K_\alpha$ diffractometer was used to confirm the phase purity. Chemical valence state of the elements and local distortions present in the samples were investigated by X-ray absorption near edge spectroscopy (XANES) and Extended x-ray absorption fine structure (EXAFS) measurements in TEY mode using hard X-ray synchrotron radiation (Indus-2, BL 9, RRCAT, Indore, India) with an estimated energy resolution (ΔE/E) about $10^{-4}$. Valence-band (VB) measurements were performed at different photon energy values in the range of 35–50 eV at the angle integrated photo emission spectroscopy (AIPES) beamline on Indus-1 synchrotron source at RRCAT, Indore, India. The instrumental resolution was estimated to be ~0.30 eV for the studied photon energy range. Prior to the photoemission measurements, the surface of the NIO bulk was scrapped in-situ using a diamond foil. For determination of the Fermi level, the Au foil was kept in an electrical contact with the sample on the same sample holder and the Fermi level was aligned using the valence-band spectrum of the Au foil. The background of valence band was corrected using Shirley method. To investigate the unoccupied states x-ray absorption near edge spectroscopy (XANES) across O K edge in TEY mode was performed at the beam line BL-01, Indus-2 at RRCAT,

Indore, India. Magnetic measurements were performed using a 7 Tesla superconducting quantum interference device (SQUID)-vibrating sample magnetometer (SVSM; Quantum Design Inc.). Electrical transport measurements were performed by the four probe method using a homemade setup.

## 3. Results and Discussions

*3.1 DFT calculations:* Energies of NIO under GGA+SO+$U$ methodology for the zigzag (ZZ), Néel, stripy and ferromagnetic configurations (*Figure 1*) are presented in *Table 1*. Our ground state of NIO corresponds to the zigzag antiferromagnetic (AFM) configuration, consistent with the previous theoretical reports [20,21,25]. The electronic structure of the ZZ magnetic configuration is calculated by employing (a) GGA, (b) GGA+$U$, (c) GGA+SO and (d) GGA+SO+$U$ methodologies, where $U$ is the electron correlation. The band dispersion around the $E_F$ and the corresponding density of states (DoS) are shown in *Figure 2(a)*. It is observed that both GGA and GGA+$U$ result in a metallic state in the absence of SOI, as also obtained in previous studies [17,20]. But, a pseudo gap of about 100 *meV* is obtained by excluding $U$ and considering only SOI (GGA+SO). Inclusion of both $U$ and SOI (GGA+SO+$U$) opens up a semiconducting gap of about 450 *meV*, that is very close to the previously reported experimental band gap of 340 *meV* [16]. The GGA+SO+$U$ electronic structure is comparable to the previous calculations using HSE06 hybrid functional [20]. This implies that for NIO it is important to consider both SO and $U$ on equal footing.

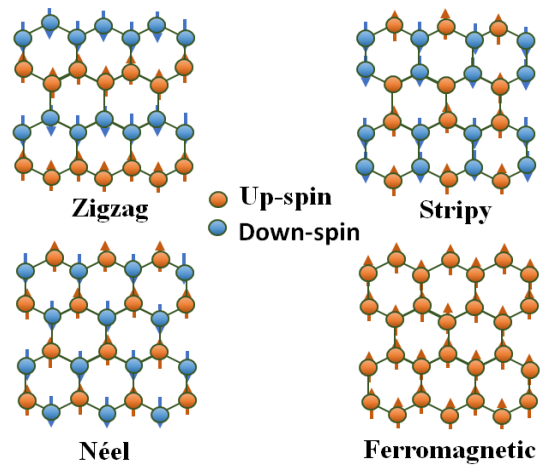

*Figure 1:* Zigzag, stripy, Néel and ferromagnetic configurations of NIO. Blue and orange circles correspond to up and down spins respectively.

*Table 1: Energies of NIO under four possible magnetic configurations calculated using GGA+SO+U method with $U_{eff}$=1.1 eV.*

| Magnetic configuration | $\Delta E$ (meV/f.u) |
| --- | --- |
| Zigzag | 0.0 |
| Stripy | 6.2 |
| Néel | 6.9 |
| Ferromagnetic | 3.0 |

In NIO, the large crystal field splitting and small on-site Coulomb repulsion due to the larger spatial extent of the 5$d$ orbital of Ir$^{4+}$ ion leads to a low-spin electron configuration, $t_{2g}^5 e_g^0$. As the $t_{2g}$ and $e_g$ orbitals are well separated due to large crystal field splitting, in previous literatures, the $j=3/2$ and $1/2$ pseudo-spin states were derived from $l_{eff}=1$ corresponding to the $t_{2g}$ states and $s=1/2$; the $e_g$ state lies much above the $j=1/2$ state. Our projected DoS (*Figure 3*) on the contrary reveal that all the bands in the vicinity of $E_F$ have admixture of $t_{2g}$ and $e_g$ orbital characters, as also reported for RuCl$_3$ [32]. Generally, the mixing of $t_{2g}$ and $e_g$ orbitals can occur either due to the SOI or due to the octahedral distortions like rotation or tilt as well as by reduction of local symmetry due to trigonal distortions, tetragonal distortions etc. In early 3$d$ TMOs like LaVO$_3$ also, a GdFeO$_3$ type of octahedral distortion aids in mixing of $t_{2g}$ and $e_g$ orbitals [33-35]. The relaxed structure from all the four methods shows distorted IrO$_6$ octahedra and the structure of GGA+SO+$U$ method is shown in *Figure 2(b)*. The octahedral distortion results in strong deviation of the $\theta_{O-Ir-O}$ angle involving two apical oxygens from the ideal 180º to 170.9º [35]. Further, the $\theta_{Ir-O-Ir}$ angle involving edge-shared oxygen is also increased from the ideal 90º to 100.7º. The octahedral distortion and mixing of $t_{2g}$ and $e_g$ orbitals are observed both in presence and absence of SOI, leading to the conclusion that the mixing is caused by local structural distortion rather than the SOI. According to our conventional wisdom, the $t_{2g}$ state is composed only of $d_{xy}$, $d_{yz}$ and $d_{zx}$ orbitals while the $e_g$ state has only $d_{x^2-y^2}$ and $d_{z^2}$ characters and a trigonal distortion can only split the $t_{2g}$ state. In NIO, since both $t_{2g}$ and $e_g$ orbitals have admixture of all the five $d$-orbitals, the local structural distortion splits both $t_{2g}$ and $e_g$ orbitals alike, resulting in five non-degenerate states respectively. Our x-ray absorption spectroscopy result presented in the following section also shows two distinct features corresponding to $e_g$ states.

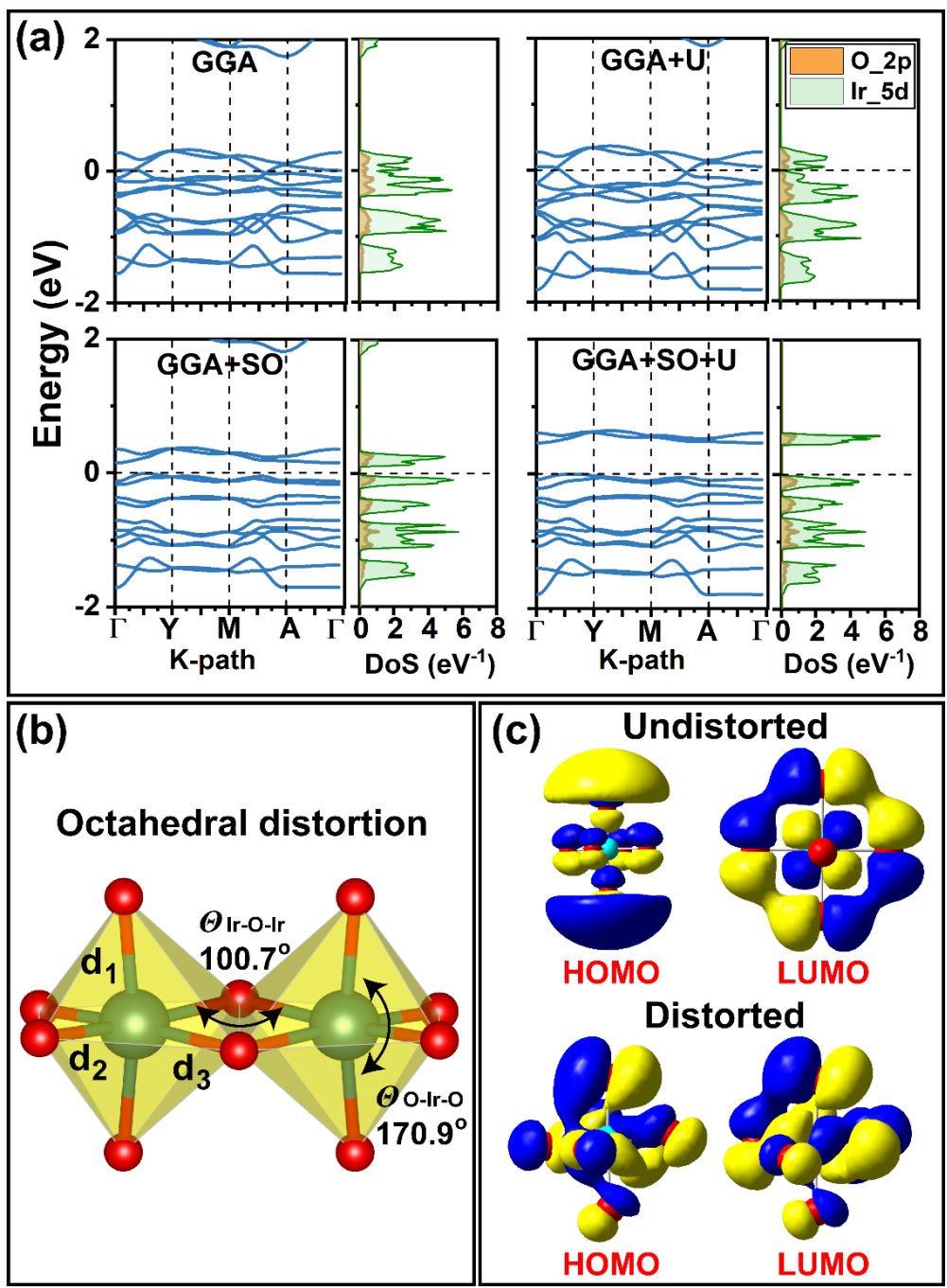

*Figure 2:* (a) Electronic structure of ZZ AFM structure using GGA, GGA+U, GGA+SO and GGA+SO+U methods (b) Local structural distortion of IrO$_6$ octahedra with iridium and oxygens represented by cyan and red spheres and (c) electron density surfaces for isodensity ($\rho$) value of 0.02 e/Å$^3$ of HOMO and LUMO for undistorted and distorted [IrO$_6$]$^{8-}$ molecular geometries.

**3.2 Quantum chemical calculations of $[IrO_6]^{8-}$ cluster:** In order to further establish the orbital mixing, we performed quantum chemical calculations of $[IrO_6]^{8-}$ molecule with spin multiplicity 2 in undistorted and distorted octahedral molecular geometries within the framework of DFT using B3LYP exchange-correlation functional, LANL2DZ basis set for iridium and 6-31g** basis set for oxygen in Gaussian 09 suite [36,37]. The equilibrium Ir-O bond lengths are around 1.88 Å in the optimized distorted octahedral geometry of $[IrO_6]^{8-}$. Hence, the bond length of 1.88 Å is chosen in the single point energy calculation of undistorted octahedral geometry also. The electron density surface corresponding to $\rho=0.02$ e/Å$^3$ for the highest occupied molecular orbital (HOMO) and lowest unoccupied molecular orbital (LUMO) for both the geometries are shown in *Figure 2(c)*. In the absence of distortion, $d_z^2$ orbital of Ir contributes to the HOMO and the LUMO is composed of $d_{xy}$ orbital, along with 2p orbitals of oxygen. But, in the distorted structure, we notice a significant mixing of $t_{2g}$ and $e_g$ orbitals of Ir with the 2p orbitals of oxygen both in the HOMO and LUMO, further confirming that local structural distortion leads to $t_{2g}$ and $e_g$ mixing.

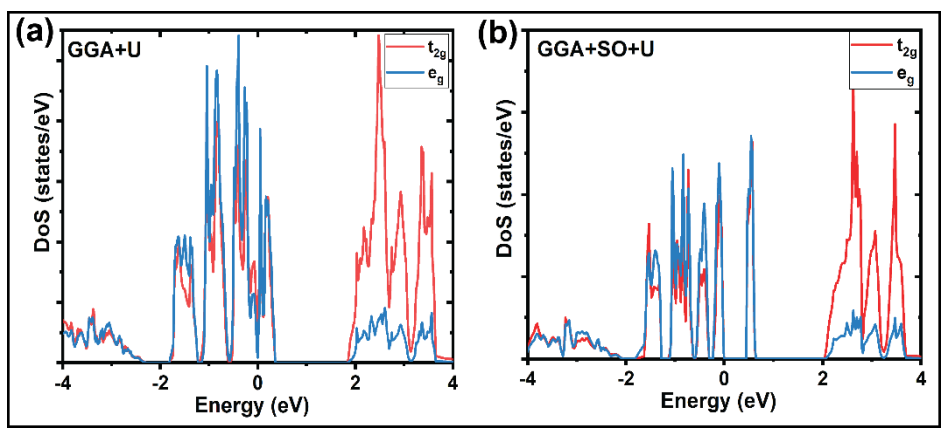

*Figure 3: Partial DoS plots for (a) GGA +U and (b) GGA+SO+U which show mixed $t_{2g}$ and $e_g$ character near the Fermi level.*

**3.3 Local distortion and $t_{2g} - e_g$ mixing:** In GGA and GGA+*U* calculated band structures three separate bands with large dispersion are observed around $E_F$. The reduction of local symmetry results in mixing of $t_{2g}$ and $e_g$ orbitals via oxygen 2p orbitals. This aids in stronger Ir 5*d* – O 2*p* hybridization which increases the overall bandwidth of Ir 5*d* orbitals and facilitates the electrons to delocalize, thereby reducing the effective strength of electron correlation. In presence of SOI, five bands are seen below the $E_F$ corresponding to *j=3/2* and *1/2* pseudo-spin states. The splitting of *j=3/2* quartet into four non-degenerate states causes the effective correlation strength to increase and leads to opening up of an energy gap. This is corroborated with reduction in octahedral distortion from $\theta_{\text{O-Ir-O}}=170°$ in GGA+$U$ to 170.9° in GGA+SO+$U$. The reduction in the distortion also alleviates the $d_{xy}$, $d_{yz}$, $d_{zx}$ mixing with $d_{x^2-y^2}$ and $d_{z^2}$ and leads to weight transfer of $e_g$ bands to higher energies, leading to reduction in bandwidth (*Figure 3*).

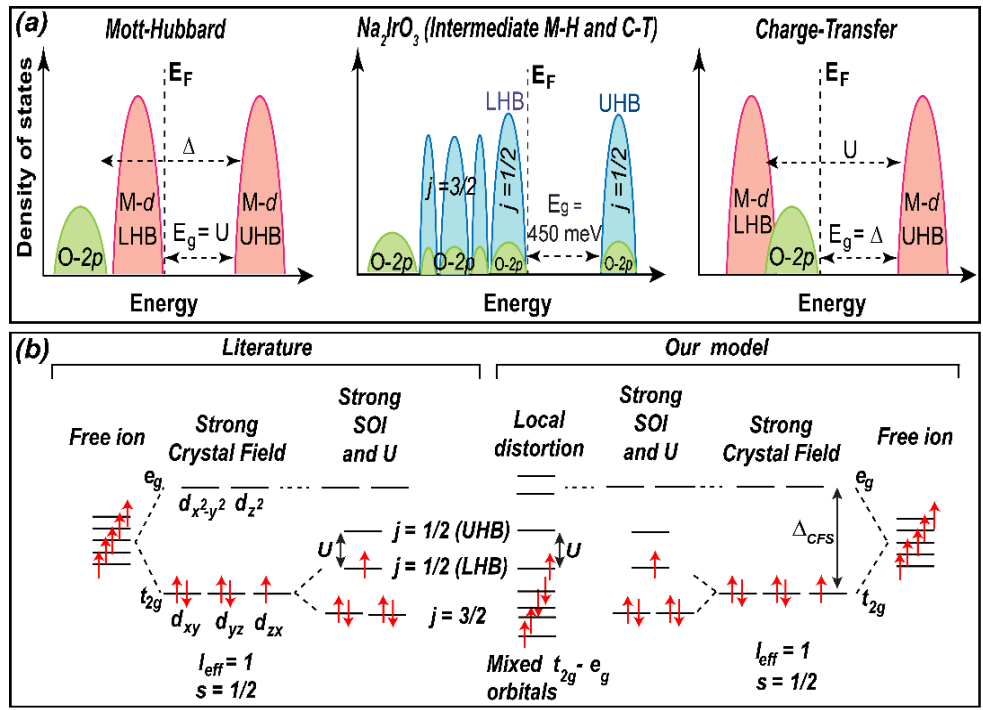

*Figure 4:* (a) Schematic of DoS of M-H and CT insulators and that of NIO in the vicinity of $E_F$ (b) Origin of j=1/2 pseudo-spin state as known previously and our proposed model.

*3.4 Mott-Hubbard insulator with charge transfer character:* A previous density functional theory (DFT) calculation using HSE06 functional showed delocalization of electrons leading to considerable decrease in the spin moment from the expected 1 $\mu_B$ for $s=1/2$ [20]. However, in presence of a strong SOI, the orbital contribution (*l*) to the total angular momentum (*j*) cannot be neglected. The *z*-component of the spin and orbital moments do not commute with the SOI operator $\widehat{H}_{SOI} = \lambda\, \hat{l} \cdot \hat{s}$ and only $j^2$, $j_z$, $l^2$ and $s^2$ are good quantum number as they commute with the SOI Hamiltonian [38]. So, we use the DFT calculated spin and orbital magnetizations values 0.35 $\mu_B$ and 0.33 $\mu_B$ to obtain spin and orbital angular momenta 0.175 and 0.33, assuming $g_s$=2.0 and $g_l$=1.0 respectively. The total angular momentum is therefore determined to be 0.5 as expected, confirming NIO to be a M-H insulator.

In a M-H or CT insulator the valence band edge has either purely metal *d*-orbital or O *2p* character respectively as illustrated in *Figure 4(a)*. Our calculated projected DoS of iridium and oxygen sites show that the valence and conduction band edges are of predominantly Ir *5d* character which is hybridized with O *2p* states (*Figure 2(a)*). We further integrate the DoS calculated using all the four methods, corresponding to iridium and oxygen ions for the *j=1/2* feature across the $E_F$ and find that the ratio of area under the curve for the two ions ($A_O/A_{Ir}$= 0.16) does not change. Interestingly, $A_O/A_{Ir}$ for the second feature below the $E_F$ corresponding to *j=3/2* state is $A_O/A_{Ir}$= 0.22, larger than the first feature, suggesting a stronger hybridization with O *2p* orbitals. Due to the presence of Ir *5d* – O *2p* hybridized features around the $E_F$, we conclude that NIO is predominantly a M-H insulator with an enhanced CT insulator character. The above results are summarized in *Figure 4(b)* that describes the energy level splitting in presence of distortion and spin orbit coupling in NIO.

*3.5 Valence band, resonant photoemission and soft x-ray absorption studies:* In order to verify these observations, we prepared phase pure bulk $Na_2IrO_3$. The phase purity of the sample is ascertained using powder x-ray diffraction. Our refined X-ray diffraction (XRD) pattern and unit cell are shown in *Figures 5(a)* and *(b)*. The peaks can be fitted with C2/m monoclinic space group and match well with previous literature. The structural parameters are comparable with our theoretically relaxed zigzag magnetic ground state structure as presented in *Table 2*. The refined structure shows three distinct bond lengths $d_1$, $d_2$ and $d_3$ respectively, 2.04 Å, 2.08 Å and 2.16 Å. We also notice a strong deviation in the experimental $\theta_{Ir-O-Ir}$ and $\theta_{O-Ir-O}$ angles as predicted by our DFT calculations. The charge state of $Ir^{4+}$ is established by performing XANES measurement as discussed in the next subsection. *Figure 6 (a)* shows the room temperature valence band spectrum (VBS) of $Na_2IrO_3$ recorded at photon energy 70 eV. The absence of spectral density or DoS at $E_F$ (*Figure 6(a)* inset) suggests insulating nature of NIO. The measured VBS is fitted with combined Gaussians and Lorentzian functions, which adequately deconvolute the whole spectrum. We observe seven features A-G in the spectra at the binding energy positions 0.9, 2.3, 4.2, 6, 7.5, 8.9 and 10.4 eV respectively (*Table 3*). To understand the nature of different features in the VBS we recorded the resonant photoemission spectrum (RPES) at room temperature with photon energies in the range 35-50 eV, which covers the Ir *5p* – Ir *5d* excitation. We normalize all the spectrum with incident beam current to remove the possibility of error due to temporal variation of incident photon flux. At the *5p→5d* photo-absorption energy (43 eV for elemental Ir) [39], besides the direct photoemission process $Ir: 5p^6\ 5d^5 + h\nu \rightarrow Ir: 5p^6\ 5d^4 + e^-$ an indirect photoemission channel is also available $Ir: 5p^6\ 5d^5 + h\nu \rightarrow [5p^5\ 5d^6]^* \rightarrow Ir: 5p^6\ 5d^4 + e^-$. As these two processes yield the same final state, by constructive interference of these two quantum mechanical processes the total photoelectron yield of 5d is enhanced. This absorption energy of various elements in the compound form is known to enhance due to correlation effects with respect to ionic state.

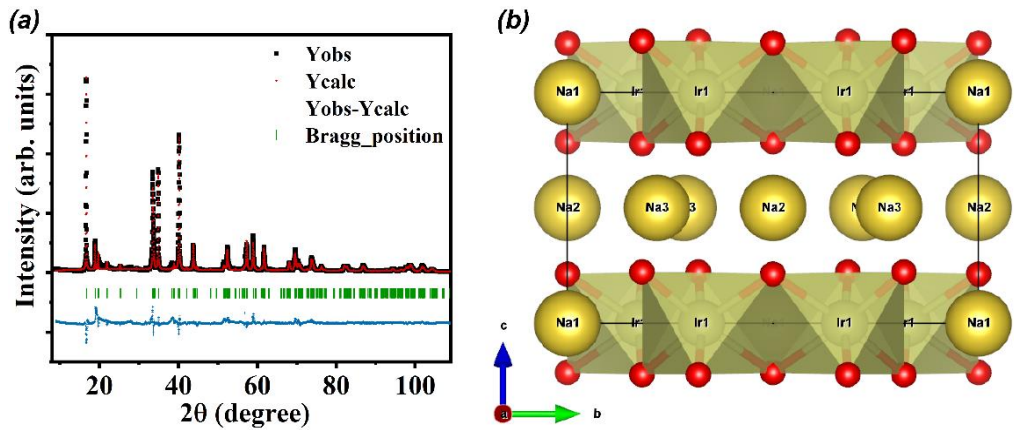

*Figure 5:* (a) Room temperature X-ray diffraction pattern and the Rietveld refinement of as synthesized NIO bulk which yields cell parameters a =5.42 Å, b=9.39 Å, c=5.62 Å and α=γ=90º, β =108.89º (b) Stacking along the c-axis in NIO as viewed from [100] direction.

*Table 2:* Structural parameters obtained from the DFT relaxed zigzag ground state and room temperature XRD data refinement.

| Structural parameters | $d_1$ (Å) | $d_2$ (Å) | $d_3$ (Å) | $\theta_{Ir-O-Ir}$ | $\theta_{O-Ir-O}$ |
|---|---|---|---|---|---|
| *Experimental* | 2.04 | 2.08 | 2.16 | 97.63º | 175.88º |
| *Calculated* | 2.06 | 2.07 | 2.08 | 100.70º | 170.9º |

*Table 3:* Assignment of all seven deconvoluted peaks.

| Feature | Binding energy (eV) | Assignment |
|---|---|---|
| A | 0.9 | Ir 5d -O 2p hybridized band. |
| B | 2.3 | Ir 5d -O 2p hybridized band. |
| C | 4.2 | O 2p Non-bonding band. |
| D | 6 | Ir 5d - O 2p hybridized band |
| E | 7.5 | Ir 5d - O 2p- Ir 6s hybridized band |
| F | 8.9 | Ir 5d - O 2p- Ir 6s hybridized band |
| G | 10.4 | Hydrocarbons |

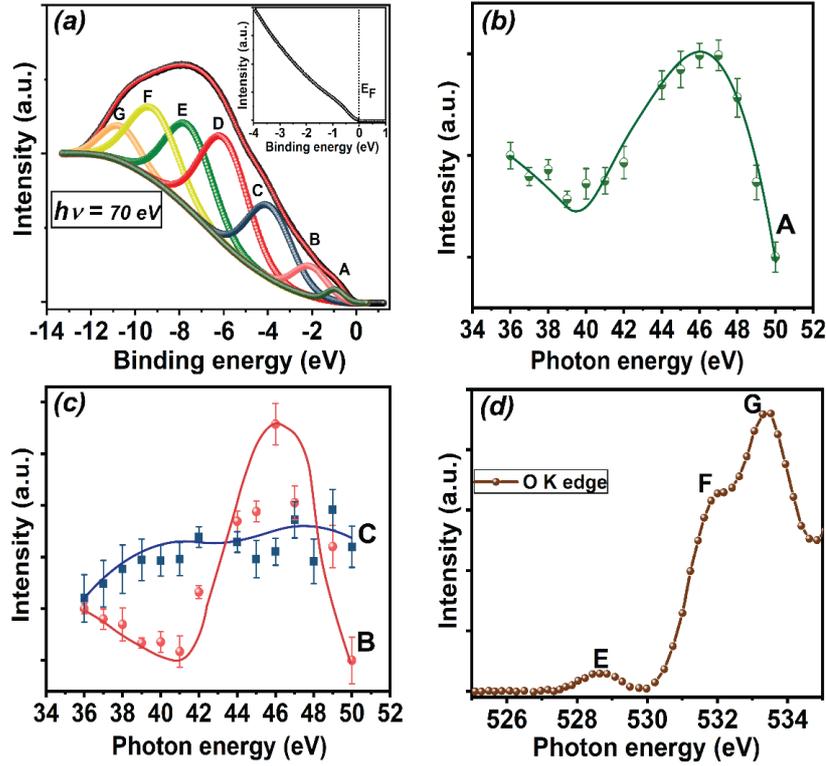

***Figure 6:*** *(a) Valence band spectrum (VBS) and the deconvoluted spectra for 70 eV photon energy (b) and (c) CIS plot of resonance photoemission spectra showing resonance and antiresonance features for features A, B and C (continuous lines are guide to the eye) (d) O K-edge of bulk NIO showing the UHB and the splitting between the $e_g$ features.*

The constant initial state (CIS) plot, which is variation of integrated intensity for any feature (obtained from area under feature) with photon energies is shown in *Figure 6(b)* and *6(c)*. Here we focus our analysis for near $E_F$ features A, B and C. It is observed that the features A and B show resonant enhancement around 46 eV, which is close to the Ir $5p \rightarrow 5d$ photo-absorption energy, confirming its Ir-$5d$ character. Besides this maximum, an antiresonance dip on lower photon energy is observed for both the features below 39 eV. Usually, the $5d^{n-1}$ final state does not show any anti-resonance dip near $5p$-$5d$ threshold, however, for $5d^n\underline{L}$ state (where $\underline{L}$ denotes hole in oxygen), an antiresonance dip on lower photon energy occurs due to increasing photoionization cross-section of oxygen ion with decrease in photon energy [39-40]. Thus features A and B are attributed to the hybridized Ir-$5d$ and O-$2p$ states with $5d^n\underline{L}$ character, which arises due to the highly covalent character of $5d$ orbital of iridium [33,41]. Feature C does not show any resonant enhancement with photon energy, revealing its O-$2p$ non-bonding character. These results further support our DFT results that NIO has characteristics of both M-H and CT insulators. A previous theoretical study had shown that the features in energy range -3 eV to $E_F$ are Ir $5d$-$t_{2g}$ and those below –3 eV are O $2p$ bands [16]. However, our experimental and theoretical results confirm the presence of O $2p$ contribution near the $E_F$.

*Figure 6(d)* shows the oxygen K-edge of NIO corresponding to the transition from core $1s$ to valence $2p$ state, which is hybridized with both Na and Ir. Here, we observe three features E, F and G at energy positions 528.8, 531.8 and 533.2 eV respectively. Previously, feature E was designated as $j=1/2$ or admixture of $j=1/2$ and $3/2$ states arising due to SO coupled $t_{2g}$ orbitals and features F and G were attributed to the split $e_g$ states due to non-local interactions [21]. It is worth noting that our calculations show that the electronic energy bands near $E_F$ have admixture of all the five d-orbitals due to non-cubic local symmetry. This local distortion caused $t_{2g}$ and $e_g$ mixing can explain the splitting observed in F and G features, since the octahedral distortion in $t_{2g}^5 e_g^0$ half-filled system which usually splits only the $t_{2g}$ states, can now split even the $e_g$ orbitals and hence we observe a doublet feature.

*3.6 Hard x-ray absorption spectroscopy:* The x-ray absorption coefficient $\mu(E)$ as a function of the incident photon energy is shown in the *Figure 7(a)*. The obtained spectra can be deconvoluted into x-ray absorption near edge spectroscopy (XANES) and extended X-ray absorption fine structure (EXAFS) regions. XANES region corresponds to excitation from $2p$ core level to unoccupied $5d$ orbital, or into the continuum where the electron is no longer associated with the atom. The spectrum of pure $IrO_2$ sample in which iridium exists in the 4+ charge state is used as reference and was simultaneously recorded with NIO. In case of the reference sample the $L_2$ and $L_3$ edges for $Ir^{4+}$ occur at 12.823 keV and 11.214 keV respectively, consistent with previous reports [43]. To enable the accurate comparison between the reference sample and NIO bulk, pre- and post-edge corrections are applied for both $L_2$ and $L_3$ edges using Athena software. It is observed that the positions of absorption edge for both NIO and standard sample $IrO_2$ exactly coincide, indicating that the iridium ion is in the 4+ valence state in NIO as well. This result, combined with the phase purity observed in our XRD pattern confirm that our bulk sample is indeed pure $Na_2IrO_3$.

EXAFS is a very useful technique to provide valuable information about the local structure of any material such as the bond length, bond angle, coordination environment etc. The *Figure 7(b) (top)* shows the modulus of Fourier transform $|\chi(R)|$ of $k^2$ weighted $\chi(k)$ signal in k-space of Ir $L_3$ edge spectra. The quantitative analysis of $|\chi(R)|$ has been carried out using model fitting with $Na_2IrO_3$ crystal structure (space group C2/m). Theoretical fitting models are created using crystallographic information obtained from Rietveld analysis of the XRD pattern. The fits are confined to R range of 1 Å < R < 2.5 Å and k range of 3 Å$^{-1}$< k < 8 Å$^{-1}$. Here within this R region, EXAFS spectra originate because of photoelectron scattering from the nearest neighbor octahedral O atoms, connected with Ir core absorber. Depending on the statistical significance, the most relevant single scatterings are taken to model theoretical EXAFS pattern. Three different Ir-O single scattering paths contribute to model fit the first shell in this certain range. Being chemically transferable, the amplitude reduction factor ($S_0^2$) is obtained from the refinement of the Ir L$_3$-edge EXAFS spectra of IrO$_2$ and is kept same for all coordinations. Similarly, same energy shift (E) is used for all co-ordinations. During the refinement cycles, the values corresponding to the average coordination distances and the mean-square relative displacement factors are refined. The modelled pattern shows good agreement with the experimentally observed behaviour with goodness of fit parameter R=0.02. The model fitting for the local structure showed three different bond lengths $d_1$, $d_2$ and $d_3$ respectively, 1.93 Å, 1.96 Å and 2.05 Å as shown in *Figure 7(b) (bottom)*.

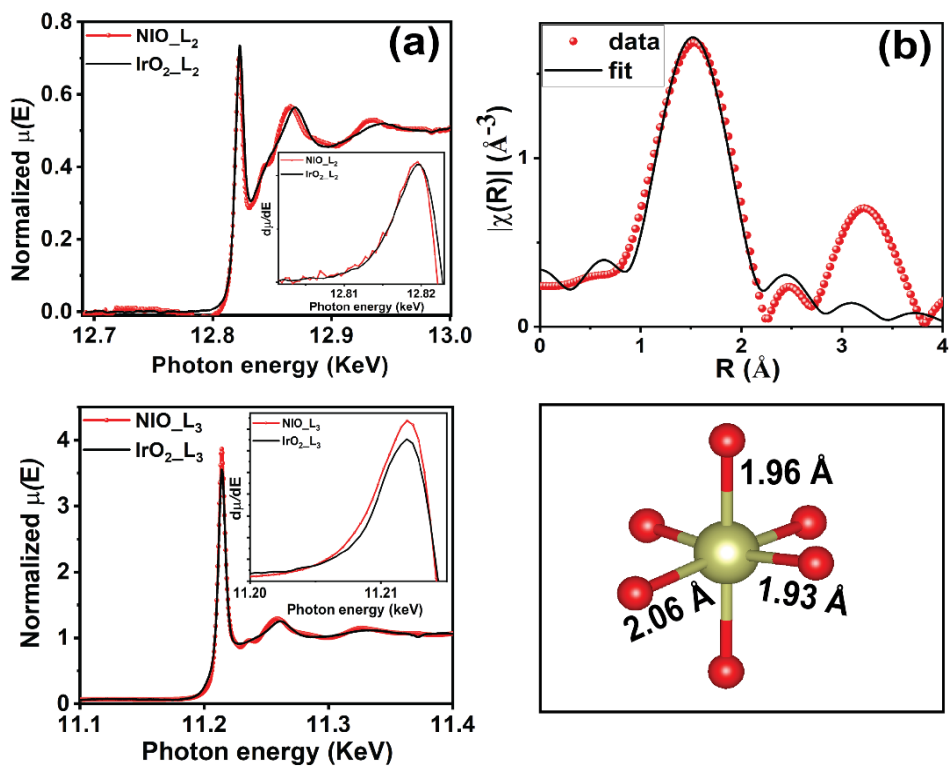

***Figure 7:*** *(a) XANES L$_2$ (top) and L$_3$ (bottom) absorption edges for Ir in IrO$_2$ and NIO bulk; insets represent first derivative of XANES. (b) Ir L$_3$ edge EXAFS analysis, (top) modulus (|χ(R)|) of the FT of k$^2$-weighted spectra [χ(k)] and (bottom) demonstrates the model fitted local octahedral Ir-O bond lengths.*

***3.7 Magnetic measurement:*** Temperature dependent magnetization behaviour of NIO bulk has been recorded in the temperature range 7K to 350K at 100 Oe magnetic field. The $\chi$ (M/H) versus T (Temperature) of NIO (*Figure 8*) shows a paramagnetic to antiferromagnetic transition ($T_N$) at 16.5 K (inset 8 (b)). We also present the $1/\chi$ vs T data linearly fitted in the region 200K to 350K (inset 8 (a)) using Curie-Weiss relation $\chi = C/(T-\theta)$, where C denotes the Curie constant and $\theta$ is the Curie-Weiss temperature. Our best fit parameters yield C=26.315 emu/g and $\theta$=-270.33 K. Deviation from a linear behaviour below 200K denotes an increasing influence of antiferromagnetic correlations at low temperatures. The large value of $\theta/T_N \sim 16.4$ denotes a high degree of frustration present in the NIO sample. Using the Curie constant $C = \frac{Ng^2\mu_B^2 j(j+1)}{3k_B}$ we also evaluate the total angular momentum *j=0.6*, corresponding to $\mu_{eff}$ =1.9 $\mu_B$ which is close enough to the theoretical value of $\mu_{eff}$ =1.74 $\mu_B$ for *j=1/2*. Similar moment values have previously been obtained in case of NIO and CaIrO$_3$ [9,44]. The higher value of *j=0.6* from the expected value could be due to the enhanced covalency or the CT character of NIO due to Ir *5d* – O *2p* hybridization around the Fermi level. We further performed transport measurement using four probe technique that shows insulating behaviour from room temperature down to 120 K (NIO is paramagnetic in this temperature range), below which the resistivity value becomes beyond the measurable range. These results clearly indicate that even in the paramagnetic state, NIO is an insulator, thus ruling out the possibility of Slater insulator behaviour.

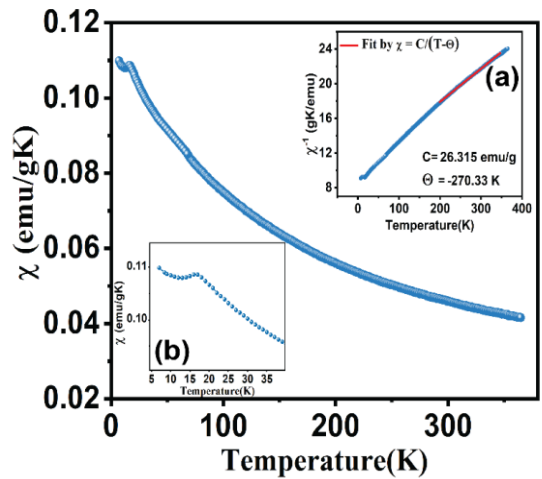

***Figure 8:*** *Magnetic susceptibility vs. temperature plot with insets showing the (a) inverse susceptibility fitted using Curie-Weiss law (red line) and (b) paramagnetic to antiferromagnetic transition at 16.5 K.*

## 4. Conclusions

We have used theoretical calculations and x-ray spectroscopy to probe the electronic structure in the vicinity of $E_F$. The origin of $j=1/2$ pseudo-spin state is governed collectively by SO interactions, electron correlation and local octahedral distortion. The $j=3/2$ and $1/2$ states contain admixture of both $t_{2g}$ and $e_g$ characters. Spin-orbit interaction further splits the energy levels within the occupied and unoccupied bands, enhancing the effective Coulomb correlation and opening a band gap. The Ir $5d$ – O $2p$ hybridization around the $E_F$ results in a M-H insulator with enhanced CT character. The absence of DoS at $E_F$ in the paramagnetic state and a localized $j=1/2$ behaviour excludes Slater insulating behaviour in NIO.

*Acknowledgments*

The authors acknowledge Dr. S.N. Jha, RRCAT for EXAFS and Dr. Rajeev Rawat, UGC-DAE CSR, Indore for resistivity measurements. PY and RR specially thank Prof. M. Kumar, SNBNCBS, Kolkata for extending computational support. MS thanks CMSD, University of Hyderabad for the computational time. We thank DST-SERB for the generous funding under the grant CRG/2019/001627. PY thanks Mr. Anupam Jana and Mr. Supriyo Majumdar for helpful discussions. We are thankful to Mr. A. Wadikar, Mr. Rakesh Sah and Mr. Sharad Karwal for their help in RPES and XAS measurements.